\documentclass[twocolumn,english,aps,prl,showpacs,superscriptaddress]{revtex4-1}
\usepackage[T1]{fontenc}
\usepackage[latin9]{inputenc}
\usepackage{geometry}
\usepackage{color}
\usepackage{babel}
\usepackage{units}
\usepackage{amsmath}
\usepackage{amssymb}
\usepackage{graphicx}
\usepackage{wasysym}
\PassOptionsToPackage{normalem}{ulem}
\usepackage{ulem}
\usepackage[unicode=true,
 bookmarks=true,bookmarksnumbered=true,bookmarksopen=true,bookmarksopenlevel=1,
 breaklinks=false,pdfborder={0 0 0},pdfborderstyle={},backref=false,colorlinks=false]
 {hyperref}
\hypersetup{pdftitle={Your Title},
 pdfauthor={Your Name},
 pdfpagelayout=OneColumn,pdfnewwindow=true,pdfstartview=XYZ,plainpages=false}

\makeatletter

\newcommand*\LyXThinSpace{\,\hspace{0pt}}

\usepackage{babel}

\usepackage{hyperref}
\def\equationautorefname~#1\null{Equation (#1)\null}

\makeatother

\begin{document}

\title{Synchronization of strongly interacting alkali-metal spins}

\author{Or Katz}
\email[Corresponding author:]{or.katz@weizmann.ac.il}

\affiliation{Department of Physics of Complex Systems, Weizmann Institute of Science,
Rehovot 76100, Israel}

\affiliation{Rafael Ltd, IL-31021 Haifa, Israel}

\author{Ofer Firstenberg}

\affiliation{Department of Physics of Complex Systems, Weizmann Institute of Science,
Rehovot 76100, Israel}
\begin{abstract}
The spins of gaseous alkali atoms are commonly assumed to oscillate
at a constant hyperfine frequency, which for many years has been used
to define the Second. Indeed, under standard experimental conditions,
the spins oscillate independently, only weakly perturbed and slowly
decaying due to random spin-spin collisions. Here we consider a different,
unexplored regime of very dense gas, where collisions, more frequent
than the hyperfine frequency, dominate the dynamics. Counter-intuitively,
we find that the hyperfine oscillations become significantly longer-lived,
and their frequency becomes dependent on the state of the ensemble,
manifesting strong nonlinear dynamics. We reveal that the nonlinearity
originates from a many-body interaction which synchronizes the electronic
spins, driving them into a single collective mode. The conditions
for experimental realizations of this regime are outlined. 
\end{abstract}
\maketitle
Binary collisions are a fundamental relaxation mechanism in atomic
spin ensembles. During a collision, a pair of atoms within the ensemble
briefly interacts, and its mutual electronic wavefunction is altered.
Since the impact parameters are random, the quantum state of the ensemble
relaxes at a rate $R$, proportional to the collisions rate $\Gamma$
\citep{Happer-book-2010}. This prevailing relaxation mechanism limits
the sensitivity of shot-noise-limited atomic sensors \citep{Kitching-sensors-review},
such as magnetometers \citep{Romalis-multipass-PRL-2013,Polzik-Magnetometery},
gyroscopes \citep{Romalis-Comagnetometer-gyro}, accelerometers \citep{Biedermann-warm-vapor-interferometer-2017},
and clocks \citep{Vanier-atomic-clocks,Campero-clocks,Kitching-clocks,Novikova-LinPerLin-Clock,Happer-end-state-clock-2004}.
It is often desirable to increase the density of the ensemble in order
to either increase the signal-to-noise ratio or to allow for miniaturization
of the device. However with the increased density, the collisional
relaxation rate $R\sim\Gamma$ increases, yielding no improvement
in the sensor sensitivity \citep{Kitching-sensors-review}.

Polarized alkali ensembles were shown to overcome this limit at low
magnetic fields \citep{happer-SERF-1977,Nonlinear-SERF,SERF_quantum_information}.
When the Zeeman splitting $\omega_{B}$ satisfies $\omega_{B}\ll\Gamma,$
the magnetic Zeeman coherences undergo a process akin to motional
narrowing via frequent spin-exchange collisions. The relaxation rate
is reduced to $R\sim\omega_{B}^{2}/\Gamma$ and so is the magnetic
linewidth. This effect, denoted as spin-exchange relaxation free (SERF),
stimulated the development of SERF magnetometers with unprecedented
sensitivities \citep{Budker-Romalis-optical-magnetometery}. 

While SERF protects the Zeeman coherences at high atomic densities,
the hyperfine coherences widely used for quantum information applications
\citep{Shu_vapor,Nunn-Raman,Polzik-RMP-2010}, radio astronomy \citep{21-cm-line}
and atomic clocks \citep{Vanier-atomic-clocks,Campero-clocks,Kitching-clocks,Novikova-LinPerLin-Clock,Happer-end-state-clock-2004}
are subject to rapid relaxation rates $R\sim\Gamma$. It is therefore
widely accepted that increased density yields faster decoherence \citep{happer-SERF-1977}.
In this letter, we prove the opposite. We derive the collisional dynamics
of a dense ensemble and find that the hyperfine coherence-time increases
significantly at high densities. We further show that rapid spin-exchange
collisions synchronize the individual spins to a single frequency,
which depends on the collective spin magnitude, leading to a unique
nonlinear many-body dynamics. 

Consider first a toy model of $N$ alkali atoms, whose ground level
encompasses an electronic spin $S=1/2$ and a nuclear spin $I=1/2$.
Most standard models describe the spin state and interactions with
an effective ensemble-averged set of equations \citep{Schwinger-average-density-matrix,Happer-Romalis-SEOP-1998,Happer-book-2010,happer-SERF-1977}.
Here we generalize these derivations and describe the many-body dynamics
of the different atoms using a general master equation formalism of
open quantum systems (see SI for the full derivation \citep{supplemnetary-1}).
The atomic state of the $n^{\mbox{th}}$ atom is described by the
observables of electronic spin $\mathbf{S}_{n}$, nuclear spin $\mathbf{I}_{n}$,
and hyperfine coherence $\mathbf{A}_{n}\equiv\mathbf{S}_{n}\times\mathbf{I}_{n}$.
 The electrons are internally coupled to their nuclei by the hyperfine
interaction $\omega_{n}\mathbf{S}_{n}\cdot\mathbf{I}_{n}$, while
every pair of electrons $\mathbf{S}_{n},\mathbf{S}_{m}$ experiences
spin-exchange interaction at a time-averaged rate $\Gamma_{mn}$.
At time scales longer than the time between collisions $\sim(\sum_{m}\Gamma_{mn})^{-1}$,
the coherences between different atoms average to zero due to the
randomness of collisions \citep{Happer-book-2010}. The many-body
dynamics of the atoms can then be represented by a compact set of
$9N$ nonlinear first-order Bloch equations,

\begin{align}
\frac{d}{dt}\left\langle \mathbf{S}_{n}\right\rangle \, & =\omega_{n}\left\langle \mathbf{A}_{n}\right\rangle +\sum_{m}\Gamma_{mn}\left(\left\langle \mathbf{S}_{m}\right\rangle -\left\langle \mathbf{S}_{n}\right\rangle \right)\label{eq:S dynamics many body-1}\\
\frac{d}{dt}\left\langle \mathbf{I}_{n}\right\rangle \, & =-\omega_{n}\left\langle \mathbf{A}_{n}\right\rangle \label{eq: I dynamics many body-1}\\
\frac{d}{dt}\left\langle \mathbf{A}_{n}\right\rangle  & =-\frac{\omega_{n}}{2}\left(\left\langle \mathbf{S}_{n}\right\rangle -\left\langle \mathbf{I}_{n}\right\rangle \right)-\sum_{m}\Gamma_{mn}\left\langle \mathbf{A}_{n}\right\rangle \label{eq:N dynamics many body I,S-1}\\
 & +\sum_{m}\Gamma_{mn}\left\langle \mathbf{S}_{m}\right\rangle \times\left\langle \mathbf{I}_{n}\right\rangle .\nonumber 
\end{align}
The first term in Eqs.\ (\ref{eq:S dynamics many body-1}-\ref{eq:N dynamics many body I,S-1})
describes the hyperfine precession of $\left\langle \mathbf{S}_{n}\right\rangle $
and $\left\langle \mathbf{I}_{n}\right\rangle $ through the coupling
with the hyperfine-coherence vector $\left\langle \mathbf{A}_{n}\right\rangle $.
The second term in Eq.\ (\ref{eq:S dynamics many body-1}) describes
the collisional exchange between the $n^{\mbox{th}}$ electronic spin
and all its neighbors. This term tends to synchronize all $\left\langle \mathbf{S}_{n}\right\rangle $
by equilibrating them with the other electronic spins $\left\langle \mathbf{S}_{m}\right\rangle $.
In Eq.~(\ref{eq:N dynamics many body I,S-1}), the second term describes
a decay of the hyperfine coherences at a rate $\sum_{m}\Gamma_{mn}$,
and the last term describes the nonlinear coherence build-up, a result
of the \emph{spin-conservative} part of the collisional interaction
\citep{supplemnetary-1}.

Spin may be exchanged between atoms when they collide, but their total
spin is conserved. Defining the atomic spin operators $\mathbf{F}_{n}=\mathbf{S}_{n}+\mathbf{I}_{n}$,
we find from Eqs.\ (\ref{eq:S dynamics many body-1}-\ref{eq:N dynamics many body I,S-1})
that the total spin of the ensemble $\left\langle \mathbf{F}\right\rangle \equiv\sum_{n}\left\langle \mathbf{F}_{n}\right\rangle $
is constant,
\begin{equation}
\frac{d}{dt}\left\langle \mathbf{F}\right\rangle \equiv\frac{1}{N}\sum_{n}\frac{d}{dt}\left\langle \mathbf{F}_{n}\right\rangle =0.\label{eq: F is conserved}
\end{equation}
In practice, this property holds for time scales shorter than the
\emph{spin destruction} rate of the ensemble (see SI \citep{supplemnetary-1}).
As no external magnetic field is included, the model is isotropic,
and the constant $\left\langle \mathbf{F}\right\rangle $ essentially
sets a preferred direction.

\begin{figure}[tb]
\begin{centering}
\includegraphics[width=8.6cm]{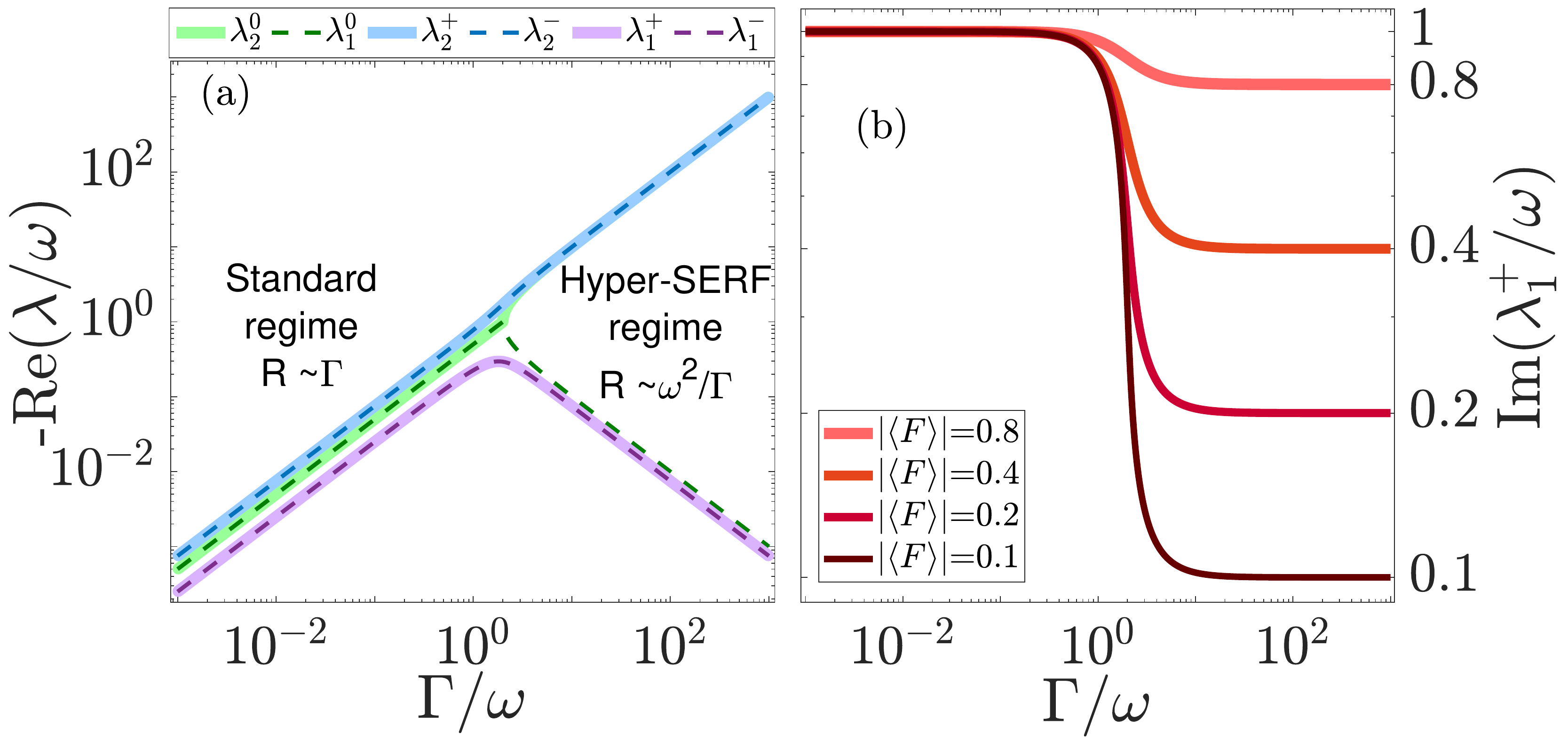}
\par\end{centering}
\caption{{\small{}(a) Relaxation rates of the hyperfine coherences for $I=1/2$
and $\left|\left\langle \mathbf{F}\right\rangle \right|=1/2$. At
high collision rates $\Gamma\gg\omega$ (high densities), the relaxation
of the $\lambda_{1}^{\pm,0}$ modes decreases. (b) Modified hyperfine
frequencies. At high collision rates, the oscillation frequency of
the hyperfine coherences becomes linearly dependent on the magnitude
of the spin $\left|\left\langle \mathbf{F}\right\rangle \right|$.\label{fig: Hyper SERF decay} }}
\end{figure}
We first consider the mean-field solution of Eqs. (\ref{eq:S dynamics many body-1}-\ref{eq:N dynamics many body I,S-1}),
assuming that $\omega_{n}\rightarrow\omega$, $\Gamma_{mn}\rightarrow\Gamma/N$,
and $\left\langle \mathbf{F}_{n}\right\rangle \rightarrow\left\langle \mathbf{F}\right\rangle $.
It follows that ${\color{blue}{\color{black}\left\langle \mathbf{S}_{n}\right\rangle =\sum_{n}\left\langle \mathbf{S}_{n}\right\rangle /N}}\equiv\left\langle \mathbf{S}\right\rangle $,
satisfying 
\begin{equation}
\ddot{\left\langle \mathbf{S}\right\rangle }+\Gamma\dot{\left\langle \mathbf{S}\right\rangle }+\omega^{2}\left(\mathbf{\left\langle S\right\rangle }-\nicefrac{1}{2}\left\langle \mathbf{F}\right\rangle \right)-\omega\Gamma\mathbf{\left\langle F\right\rangle \times\left\langle S\right\rangle }=0.\label{eq: S mean field dynamics}
\end{equation}
Since $\left\langle \mathbf{F}\right\rangle $ is constant, Eq.\ (\ref{eq: S mean field dynamics})
is a set of three linear non-homogeneous equations, whose general
solution is
\[
\left\langle S_{q}\right\rangle =\frac{1}{2}\left\langle F_{q}\right\rangle +\sum_{i=1}^{2}a_{i}^{q}e^{-\lambda_{i}^{q}t}.
\]
Here, the subscript $q=0,\pm$ denotes the three directions $\hat{z,}\left(\hat{x}\pm i\hat{y}\right)/\sqrt{2}$,
with the $\hat{z}$ axis defined as the direction of the vector $\left\langle \mathbf{F}\right\rangle $,
and the six coefficients $a_{i}^{q}$ determine the weights of the
modes and depend on the initial condition of the spins. The time-dependent
dynamics are described by six complex eigenvalues
\begin{equation}
\left(\begin{array}{c}
\lambda_{1,2}^{0}\\
\lambda_{1,2}^{+}\\
\lambda_{1,2}^{-}
\end{array}\right)=\frac{1}{2}\left(\begin{array}{c}
-\Gamma\pm\sqrt{\Gamma^{2}-4\omega^{2}}\\
-\Gamma\pm\sqrt{\Gamma^{2}+4i\Gamma\omega\left|\left\langle \mathbf{F}\right\rangle \right|-4\omega^{2}}\\
-\Gamma\pm\sqrt{\Gamma^{2}-4i\Gamma\omega\left|\left\langle \mathbf{F}\right\rangle \right|-4\omega^{2}}
\end{array}\right),\label{eq:full eigenvalues}
\end{equation}
where $\lambda_{1,2}^{0},$ $\lambda_{1,2}^{+}$, and $\lambda_{1,2}^{-}$
are the eigenvalues of $\left\langle S_{0}\right\rangle $, $\left\langle S_{+}\right\rangle $,
and $\left\langle S_{-}\right\rangle $ respectively. The real part
of these eigenvalues, associated with the relaxation rate $R$, is
shown in F\textcolor{black}{ig.\ \ref{fig: Hyper SERF decay}(a)
for a partially polarized ensemble $\left|\left\langle \mathbf{F}\right\rangle \right|=1/2$.}

In standard hot vapor experiments, the alkali densities are kept low,
such that $\Gamma\ll\omega$. In this regime, the eigenvalues in (\ref{eq:full eigenvalues})
are approximately given by
\begin{equation}
\left(\begin{array}{c}
\lambda_{1,2}^{0}\\
\lambda_{1}^{\pm}\\
\lambda_{2}^{\pm}
\end{array}\right)\approx\left(\begin{array}{c}
\pm i\omega-\Gamma/2\\
\pm i\omega-\left(1-\left|\left\langle \mathbf{F}\right\rangle \right|\right)\Gamma/2\\
\pm i\omega-\left(1+\left|\left\langle \mathbf{F}\right\rangle \right|\right)\Gamma/2
\end{array}\right).\label{eq:low_density}
\end{equation}
The oscillation frequency of the hyperfine coherences $\left|\mbox{Im}\left(\lambda\right)\right|=\omega$
is constant. The relaxation rate of the $\lambda_{1,2}^{0}$ modes,
associated with the so-called clock transition ($0-0$) used by atomic
frequency standards \citep{Vanier-atomic-clocks,Kitching-clocks,Novikova-LinPerLin-Clock},
is \textcolor{black}{$R=\Gamma/2$. The end resonances relax at $R=\left(1-\left|\left\langle \mathbf{F}\right\rangle \right|\right)\Gamma/2$,
leading to the well-known line narrowing for $\left|\left\langle \mathbf{F}\right\rangle \right|\rightarrow1$
\citep{Happer-end-state-clock-2004}.}

\begin{figure}[tb]
\begin{centering}
\includegraphics[viewport=0bp 0bp 960bp 445bp,width=8.6cm]{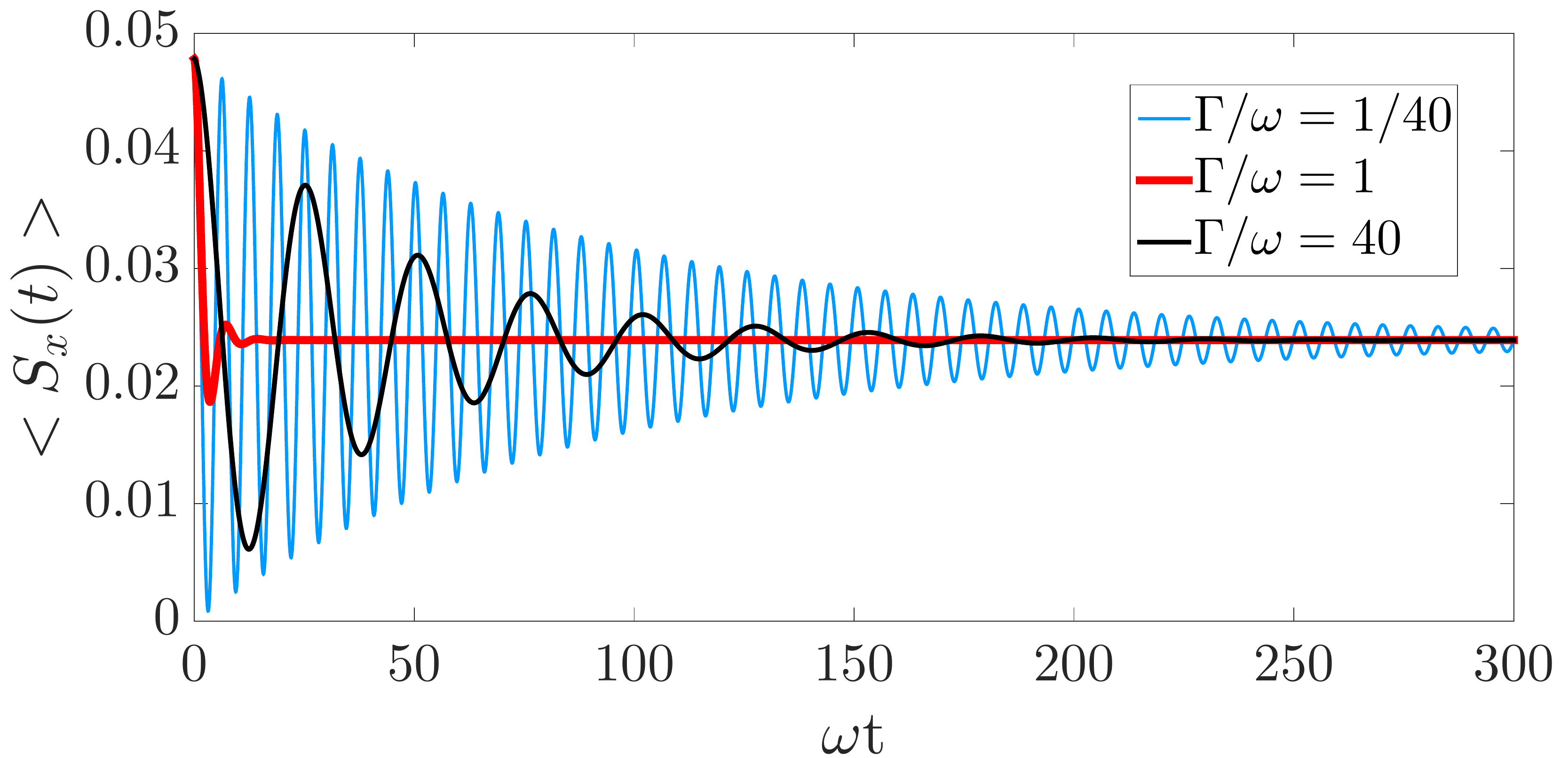}
\par\end{centering}
\caption{{\small{}Numerical simulation of the mean-field case. The coherence
time revives at high densities $\Gamma\gg\omega$. \label{fig: numerical mean field}}}
\end{figure}
 In the strong interaction regime $\Gamma\gg\omega$, the hyperfine
oscillation is strongly perturbed by spin-exchange collisions, and
the eigenvalues in (\ref{eq:full eigenvalues}) become

\begin{equation}
\left(\begin{array}{c}
\lambda_{2}^{0}\\
\lambda_{1}^{0}\\
\lambda_{2}^{\pm}\\
\lambda_{1}^{\pm}
\end{array}\right)\approx\left(\begin{array}{c}
-\Gamma\\
-\omega^{2}/\Gamma\\
\pm i\omega\left|\left\langle \mathbf{F}\right\rangle \right|-\Gamma\\
\pm i\omega\left|\left\langle \mathbf{F}\right\rangle \right|-\left(1-\left|\left\langle \mathbf{F}\right\rangle \right|^{2}\right)\omega^{2}/\Gamma
\end{array}\right).\label{eq:H-SERF eigenvalues}
\end{equation}
We find that the relaxation of the $\lambda_{1}^{\pm,0}$ modes scales
as $\omega^{2}/\Gamma$, which we attribute to motional narrowing;
increasing the collision rate $\Gamma$ slows down the hyperfine decoherence.
We denote this property as hyper-SERF, as the hyperfine coherences
become free from spin-exchange relaxation. Furthermore and quite uniquely,
the hyperfine frequency becomes dependent on the absolute magnitude
of the spin $\left|\left\langle \mathbf{F}\right\rangle \right|$.
The modified frequency of the $\lambda_{1,2}^{\pm}$ modes, shown
in Fig.\ \ref{fig: Hyper SERF decay}(b), is given by $\omega\left|\left\langle \mathbf{F}\right\rangle \right|$.
On the other hand, the $\lambda_{1,2}^{0}$ modes have no oscillatory
terms, indicating that the $0-0$ clock-transition will ``stop ticking''.

\begin{figure}[t]
\begin{centering}
\includegraphics[width=8.6cm]{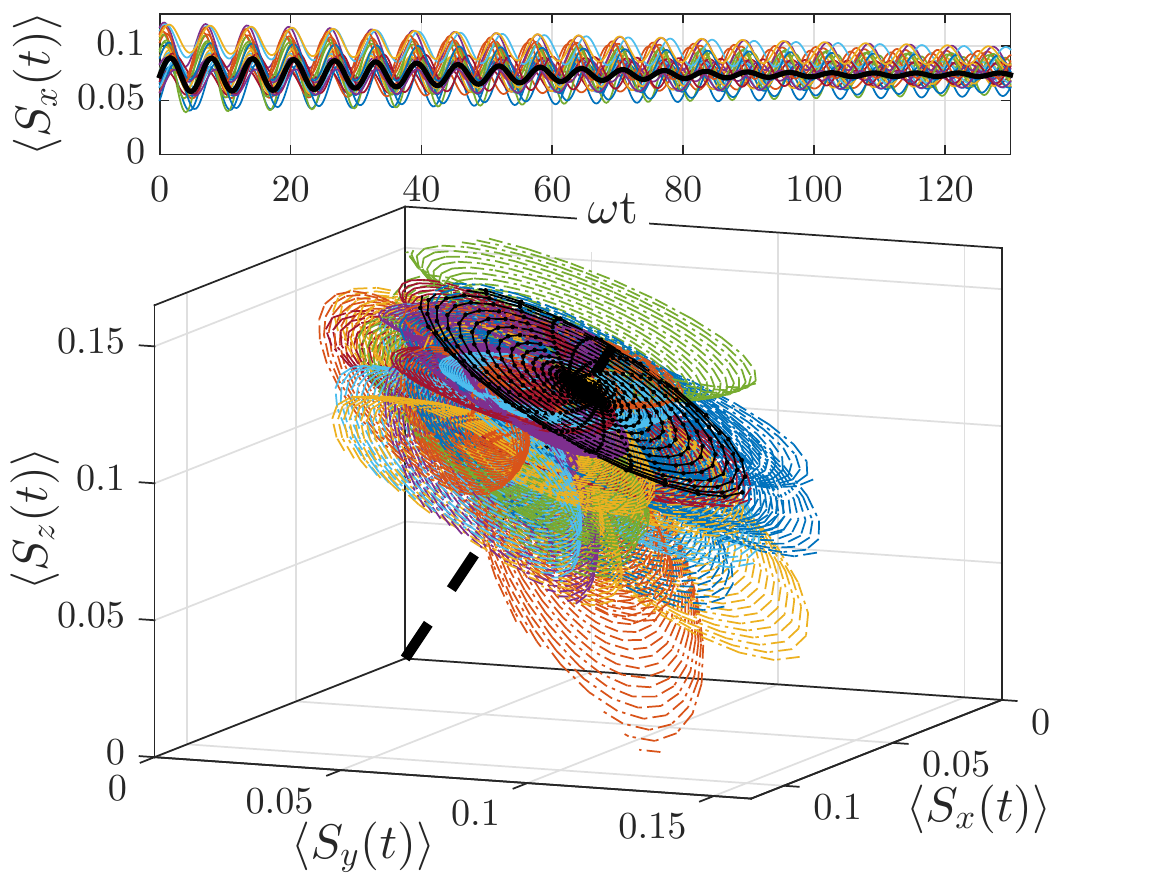}
\par\end{centering}
\caption{{\small{}Precession of the electronic spins in the standard, low density
regime with $\Gamma=\omega/100$ (25 out of N=100 simulated spins
are shown). Each electronic spin $\left\langle \mathbf{S}_{n}\right\rangle $
precesses independently around its local vector $\left\langle \mathbf{F}_{n}\right\rangle $,
slowly decaying due to collisions. The mean electronic spin (black)
precesses around the conserved spin $\left\langle \mathbf{F}\right\rangle $
(black dotted line), dephasing at an increased rate $R\sim\mbox{max}\left(\Gamma,\Delta\omega\right)$.
\label{fig: spirals G<<w}}}
\end{figure}

To understand the nature of this mechanism, we generalize the mean-field
result by numerically solving Eqs.~(\ref{eq:S dynamics many body-1}-\ref{eq:N dynamics many body I,S-1})
and obtaining the many-body dynamics of the spins. The initial values
of $\left\langle \mathbf{S}_{n}\right\rangle ,\left\langle \mathbf{I}_{n}\right\rangle ,\left\langle \mathbf{A}_{n}\right\rangle $
are derived from the initial density matrices of the atoms $\rho_{n}$.
We start with an optically pumped vapor in a spin-temperature distribution
$\tilde{\rho}_{n}=\mbox{exp}(-\beta F_{n}^{z})/Z$, where $0\leq\beta\leq1$
determines the degree of polarization, and $Z$ is a normalization
factor \citep{Happer-Romalis-SEOP-1998}. To generate initial hyperfine
coherences, we perturb $\tilde{\rho}_{n}$ by tilting the electronic
spins by angles $\theta_{n}^{y},\theta_{n}^{z}$ and the nuclear spins
by angles $\phi_{n}^{y},\phi_{n}^{z}$, such that $\rho_{n}=U_{n}\tilde{\rho}_{n}U_{n}^{\dagger}$
with the rotation matrices
\[
U_{n}=e^{i\theta_{n}^{z}S_{z}}e^{i\theta_{n}^{y}S_{y}}e^{i\phi_{n}^{z}I_{z}}e^{i\phi_{n}^{y}I_{y}}.
\]
We first simulate the mean-field solution for $N=100$, $\omega_{n}=\omega$,
and $\Gamma_{mn}=\Gamma/N$,{\small{} }as shown in Fig. \ref{fig: numerical mean field}{\small{}.
}The initial conditions are given by $\theta_{n}^{z}=\phi_{n}^{y}=\phi_{n}^{z}=0$,
$\theta_{n}^{y}=\pi/8$ and $\beta=0.51$ ($\left|\left\langle \mathbf{F}\right\rangle \right|=\nicefrac{1}{4}$).
We find indeed that the coherence time of the mean spin $\left\langle S_{x}\right\rangle $
is improved at high collision rate $\Gamma$. We further simulate
the many-body dynamics of the spins for unequal initial values and
unequal interaction strengths $\omega_{n}$ and $\Gamma_{mn}$. We
set $\beta=0.73$ ($\left|\left\langle F\right\rangle \right|=0.32$),
$\theta_{n}^{y},\theta_{n}^{z}$$\sim\mathcal{N}\left(\nicefrac{\pi}{3},\nicefrac{\pi}{15}\right)$
,$\phi_{n}^{y},\phi_{n}^{z}\sim\mathcal{N}\left(\nicefrac{\pi}{6},\nicefrac{\pi}{30}\right)$,
randomly sampled from a normal distribution $\mathcal{N}\left(\mu,\sigma\right)$
with mean $\mu$ and standard deviation $\sigma$, resulting with
unequal initial spin orientations. The collision rates $\Gamma_{mn}=\Gamma p_{mn}$
are set by generating a random double stochastic matrix $p_{mn}$.
For the generality of the model, we also allow a spread for the atomic
hyperfine frequencies $\omega_{n}\sim\mathcal{N}\left(\omega,\omega/50\right)$.
In the standard, low density, regime ($\Gamma\ll\omega_{n}$), the
individual electronic spins preces independently at their inherent
frequencies $\omega_{n}$, forming spiral trajectories around their
local spin vectors $\left\langle \mathbf{F}_{n}\right\rangle $, as
shown in Fig.~\ref{fig: spirals G<<w}. The local spin vectors slowly
relax to their equilibrium state $\left\langle \mathbf{F}_{n}\right\rangle \rightarrow\left\langle \mathbf{F}\right\rangle =\frac{1}{N}\sum_{n}\left\langle \mathbf{F}_{n}\right\rangle $
due to spin-exchange collisions, at a rate $R\sim\Gamma/2$. As a
result, the spin coherences decay, and the center of each spiral adiabatically
follows $\left\langle \mathbf{F}_{n}\right\rangle $. The mean electronic
spin $\frac{1}{N}\sum_{n}\left\langle S_{x}^{n}\right\rangle $ (black
line in Fig. \ref{fig: spirals G<<w}), decays faster than the individual
spins $\left\langle S_{x}^{n}\right\rangle $. This results from an
additional (inhomogeneous) dephasing of the different hyperfine frequencies
$\omega_{n}$ with a relaxation rate $R\sim[\sum_{n}(\omega_{n}-\omega)^{2}]^{1/2}\equiv\Delta\omega$.
\begin{figure}[t]
\begin{centering}
\includegraphics[width=8.6cm]{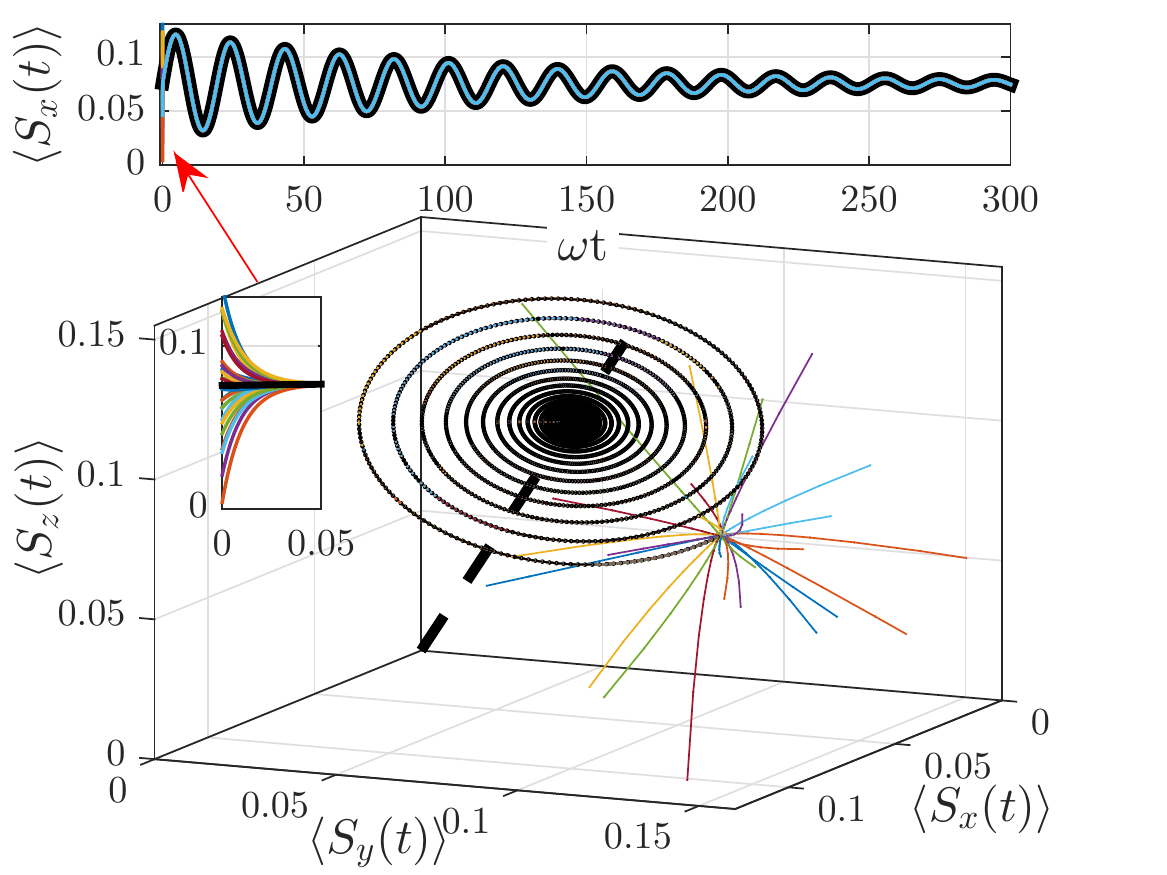}
\par\end{centering}
\caption{{\small{}Synchronization of the electronic spins in the strong interaction
regime with $\Gamma=100\omega$ (25 out of N=100 simulated spins are
shown). The electronic spins synchronize rapidly after $t\sim\Gamma^{-1}$
to a common electronic mode (black). The electronic spins precess
coherently at a modified, spin-dependent, frequency $\Omega$ and
decay at a slow rate $R\sim\omega^{2}/\Gamma$. \label{fig: synchronization1} }}
\end{figure}

In the strong interaction regime ($\Gamma\gg\omega$), the electronic
spins no longer precess individually, but rather synchronize to a
single trajectory as shown in Fig.~\ref{fig: synchronization1}.
All spins precesses around the mean spin $\left\langle \mathbf{F}\right\rangle $
with identical frequency of oscillation $\Omega$. The synchronization
time is rapid, scaling as $\Gamma^{-1}$. To reveal the synchronization
mechanism, we expand Eqs.$\;$(\ref{eq:S dynamics many body-1}-\ref{eq:N dynamics many body I,S-1})
by the small parameter $\omega/\Gamma$, keeping only second order
terms \citep{supplemnetary-1},
\begin{align}
\frac{d}{dt}\left\langle \mathbf{F}_{n}\right\rangle  & =\sum_{m}\Gamma_{mn}\left(\left\langle \mathbf{S}_{m}\right\rangle -\left\langle \mathbf{S}_{n}\right\rangle \right)\label{eq: Semi conservation model}\\
\frac{d}{dt}\left\langle \mathbf{S}_{n}\right\rangle  & \approx\sum_{m}\Gamma_{mn}\left(\left\langle \mathbf{S}_{m}\right\rangle -\left\langle \mathbf{S}_{n}\right\rangle \right)+\omega_{n}\left\langle \mathbf{F}_{n}\right\rangle \times\left\langle \mathbf{S}_{n}\right\rangle \nonumber \\
- & \frac{\omega_{n}^{2}}{\Gamma_{n}}\left(\left\langle \mathbf{S}_{n}\right\rangle -\nicefrac{1}{2}\left\langle \mathbf{F}_{n}\right\rangle \right)\label{eq:semi tops}
\end{align}
\textcolor{black}{This set of equations is known as the ``tops model''
\citep{Ritort-Tops-Model-1998},} with $\omega_{n}\left\langle \mathbf{F}_{n}\right\rangle $
playing\textcolor{black}{{} the role of a loc}al external torque. The
first term in Eq.$\;$(\ref{eq:semi tops}) initially dominates and
synchronizes the electronic spins over a transient time $\sim\Gamma^{-1}$,
as shown in Fig.\ \ref{fig: synchronization1}. Once the electronic
spins are synchronized $\left\langle \mathbf{S}_{m}\left(t\right)\right\rangle \approx\left\langle \mathbf{S}_{n}\left(t\right)\right\rangle $,
the spin vectors $\left\langle \mathbf{F}_{n}\right\rangle $ remain
approximately constant {[}Eq.\ (\ref{eq: Semi conservation model}){]}.
The second term in Eq.\ (\ref{eq:semi tops}) describes a local torque
exerted on $\left\langle \mathbf{S}_{n}\right\rangle $ by the local
field $\omega_{n}\left\langle \mathbf{F}_{n}\right\rangle $. We note
that the directions and magnitudes of these local fields could be
random. The third and least dominant term in Eq.\ (\ref{eq:semi tops})
describes the slow relaxation of the electronic spin $\left\langle \mathbf{S}_{n}\right\rangle $
towards its steady value $\left\langle \mathbf{F}_{n}\right\rangle /2$
at the hyper-SERF rate $\omega_{n}^{2}/\Gamma_{n}$. It is interesting
to note that, although the electronic spins are frustrated by the
different local fields $\omega_{n}\left\langle \mathbf{F}_{n}\right\rangle ,$
the synchronization term overcomes this frustration in the strong-interaction
regime. As a result, the synchronized electronic spins precess collectively
around an effective mean field
\begin{equation}
\mathbf{\boldsymbol{\Omega}}\approx\frac{1}{N}\sum_{n}\omega_{n}\left\langle \mathbf{F}_{n}\right\rangle .\label{eq:mean field precession}
\end{equation}
Hence electronic spins with random initial orientations are phase-synchronized,
and consequently precess coherently around the vector $\boldsymbol{\mathbf{\Omega}}$,
with a new collective modified hyperfine frequency $\Omega$. Note
that our result are valid also for the case of nonequal frequencies
$\omega_{n}$. This frequency depends on the polarization of the spin
vectors $\left\langle \mathbf{F}_{n}\right\rangle $, recovering the
mean-field results when $\omega_{n}=\omega$. Since the vectors $\left\langle \mathbf{F}_{n}\right\rangle $
do not synchronize, the directions of the nuclear spins $\left\langle \mathbf{I}_{n}\right\rangle $
remain unsynchronized as well. Nevertheless, the different nuclear
spins precess coherently, experiencing the slow electronic relaxation
$\omega_{n}^{2}/\Gamma_{n}$.

It is also instructive to interpret our results from the viewpoint
of collision-driven thermal equilibration, by extending the description
of the SERF effect in Ref. \citep{happer-SERF-1977} and considering
the hyperfine interaction as an out-of-equilibrium term.  At \textit{low}
atomic densities, spin-exchange collisions reduce the electron-nuclear
coherence, as they redistribute the electronic spin between different
atoms. At the same time, the hyperfine interaction strongly couples
the nuclear spin to the electron spin within each atom. Consequently,
the system is driven into a so-called spin-temperature distribution
$\rho_{n}=\mbox{exp}(-\vec{\beta}\mathbf{F}_{n})/Z$ with no hyperfine
coherence, thus maximizing the entropy of the spin degrees-of-freedom
\citep{Anderson-Pipkin,spin-temperature2}. The mean thermalization
rates of the different hyperfine coherences correspond to the decay
rates of Eq.~(\ref{eq:low_density}) (proportional to $\Gamma$).
In contrast, at \textit{high} atomic densities, the electron spins
alone quickly thermalize (at a rate $\Gamma$) into a spin-temperature
distribution $\rho_{n}^{s}=\mbox{exp}(-\vec{\beta}_{s}\mathbf{S}_{n})/Z_{s}$
through the spin-synchronizing term in Eq.~(\ref{eq:S dynamics many body-1}).
This thermalization leads to rapid loss of any initial correlations
between the electronic and nuclear spins, making the electronic spins
act as a single macroscopic magnetic moment on the nuclear spins $\left\langle \mathbf{A}_{n}\right\rangle \approx\left\langle \mathbf{S}\right\rangle \times\left\langle \mathbf{I}_{n}\right\rangle $.
Application of this result to Eq.~(\ref{eq:S dynamics many body-1})
shows that $|\vec{\beta}_{s}|=2\text{atanh}\left(2\left|\left\langle \mathbf{S}\right\rangle \right|\right)$
is constant in magnitude but precesses according to $\partial_{t}\hat{\beta}_{s}=\hat{\beta}_{s}\times\boldsymbol{\mathbf{\Omega}}$,
i.e., the electronic spins oscillate around the modified hyperfine
vector $\boldsymbol{\mathbf{\Omega}}$. In turn, the nuclear spins
precess around the electronic spin $\left\langle \mathbf{S}\right\rangle $
as suggested by Eq.~(\ref{eq: I dynamics many body-1}), also with
a precession frequency $\mathbf{\Omega}$. Full thermalization of
the nuclear spins happens slowly, at an approximate rate $\sim\Gamma\left(\omega/\Gamma\right)^{2}$,
where $\left(\omega/\Gamma\right)^{2}$ is the small angular loss
during the synchronization time, similar to the loss in standard SERF
of the Zeeman coherences \citep{happer-SERF-1977}.

Our model predicts several new physical phenomena in the strong interaction
regime $\Gamma\gg\omega$. The first prediction is the motional narrowing
of the hyperfine coherence, leading to its slow relaxation with a
rate that scales as $\omega^{2}/\Gamma$ rather than $\Gamma$. The
second prediction of the model is the nonlinear splitting of the hyperfine
levels, ``dressed'' by the collisional interaction, such that both
electronic and nuclear spins should precess at a rate $\omega\left\langle \left|\boldsymbol{F}\right|\right\rangle $.
The splitting depends linearly on the magnitude of the spin, and should
therefore vary for different optical-pumping rates. This dependence
can thus lead to intriguing nonlinear behavior when the probing scheme
inherently involves optical pumping, such as in coherent population
trapping (CPT) \citep{cpt}. A third prediction pertains to the case
of nonzero bandwidth $\Delta\omega$. For alkali ensembles, a mixture
of different species with different hyperfine frequencies $\omega_{n}$
effectively features nonzero $\Delta\omega$. In these hybrid ensembles,
the electronic spins of all species would synchronize and oscillate
in a common mode. The synchronization mechanism can be optically probed
by measuring the oscillation frequency of each specie separately \citep{Hybrid_spin_exchange}.\textcolor{black}{{}
}
\begin{figure}[tb]
\begin{centering}
\includegraphics[width=8.6cm]{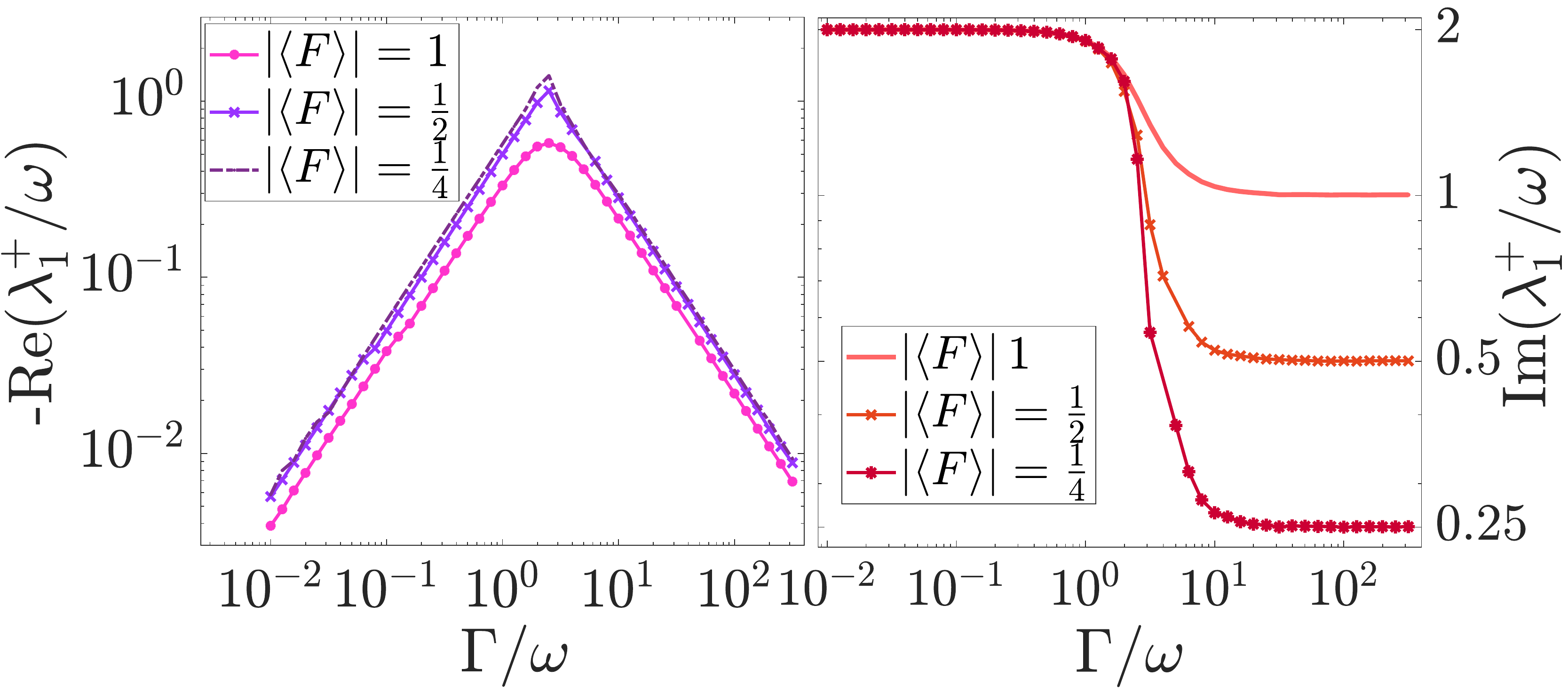}
\par\end{centering}
\caption{{\small{}Numerical calculation of hyper-SERF for $I=3/2$, }for different
initial polarizations $\left|\left\langle \mathbf{F}\right\rangle \right|${\small{}.
Shown are the dominant relaxation rate of $\left\langle S_{x}\right\rangle $
(left) and its frequency (right). The results are qualitatively similar
to the $I=1/2$ case (note that here the maximal spin is $\left|\left\langle \mathbf{F}\right\rangle \right|=2$).
\label{fig: Hyper SERF I=00003D3/2} }}
\end{figure}

We analyzed above a toy model with $I=1/2$ and no magnetic field
($\vec{B}=0$). To verify that the hyper-SERF features persist for
$I>1/2$ we numerically solved the master equation \citep{supplemnetary-1}.
Fig.~\ref{fig: Hyper SERF I=00003D3/2} presents the dominant relaxation
rate and frequency of $\left\langle S_{x}\right\rangle $ for atoms
with $I=3/2$, initialized with $\theta^{z}=\phi^{y}=\phi^{z}=0$,
$\theta^{y}=\pi/8$. These results show that the toy model results
are qualitatively valid for $I>1/2$ spins. If a magnetic field is
applied, both the direction and magnitude of $\left\langle \mathbf{F}\right\rangle $
could vary in the presence of collisions. At magnetic fields $B\lesssim10\,\text{Gauss}$,
the Zeeman splitting is small $(g_{s}B\ll\omega$, where $g_{s}$
is the gyromagnetic ratio), $\left\langle \mathbf{F}\right\rangle $
slowly precesses around $\vec{B}$, and our solution for the hyperfine
coherences adiabatically follows the instantaneous $\left\langle \mathbf{F}\right\rangle $.

\textit{Experimental Roadmap. }Hyper-SERF with $I=3/2$ can be experimentally
realized using $\mbox{\ensuremath{^{41}}K}$, which has the lowest
hyperfine frequency $2\omega_{{\scriptscriptstyle \textsc{\textnormal{K}}}}\sim254\,(2\pi)$
MHz (the factor of 2 enters since $I=3/2$). The density required
for entering the strong interaction regime is $n_{{\scriptscriptstyle \textsc{\textnormal{K}}}}>\omega_{{\scriptscriptstyle \textsc{\textnormal{K}}}}/(\sigma_{{\scriptscriptstyle \textsc{\textnormal{SE}}}}\bar{v})\approx5\cdot10^{17}\;\mbox{\ensuremath{\mathrm{cm}^{-3}}}$,
where $\bar{v}\approx10^{5}\;\mbox{cm/sec}$ is the mean thermal velocity
at $T\approx600\;\text{\ensuremath{^{\circ}}C}$ and $\sigma_{{\scriptscriptstyle \textsc{\textnormal{SE}}}}=1.5\cdot10^{-14}\;\mathrm{cm}^{2}$
is the spin-exchange cross-section. High temperature cells based on
sapphire windows were demonstrated \citep{Cundiff-Hot-vapor-cell},
as sapphire can withstand alkali metal at elevated temperatures for
long time.

To observe hyper-SERF dynamics, relaxation mechanisms of the vapor
should be kept low with respect to the hyperfine frequency. We propose
to utilize a miniature cell of length $L=100\;\text{\ensuremath{\mu}m}$
with $1\;\text{amg}$ of $\text{\ensuremath{N_{2}}}$ buffer gas at
$T=620\;\text{\ensuremath{^{\circ}}C}$ (corresponding to $n_{{\scriptscriptstyle \textsc{\textnormal{K}}}}=1.7\cdot10^{18}\;\mbox{\ensuremath{\mathrm{cm}^{-3}}}$
and $R_{\text{SE}}=2.5\cdot10^{9}\;\text{se\ensuremath{c^{-1}}}$).
Estimation of the main relaxation mechanisms of the vapor based on
the theory in Refs.~\citep{Walker-2000-relaxations-singlet,Walker-2000-spin-destruction-triplet}
yields $R_{\text{SD}}<5\cdot10^{6}\;\text{se\ensuremath{c^{-1}}}$
(see SI \citep{supplemnetary-1}), so that spin exchange dominates.
The $\text{\ensuremath{N_{2}}}$ buffer gas can mitigate both the
interaction with the walls and other molecular relaxations. Choosing
$\text{\ensuremath{N_{2}}}$ also enables efficient optical pumping
at elevated densities, by quenching excited-state alkali atoms and,
consequently, avoiding spontaneous emission of stray photons \citep{radiation-trapping}.
An effective optical-depth of $\sim700$ is expected, with an optical
linewidth of $\sim70\;\text{GHz}$ dominated by alkali self-broadening
\citep{self-broadening} and pressure broadening. At these conditions
the probability to spontaneously radiate a photon is kept low $(\sim0.2\%)$,
and the photon-multiplicity is moderate $(\sim30)$, mitigating radiation
trapping \citep{radiation-trapping}. Optical pumping at a rate of
up to $R_{P}\approx1\;\text{GHz}$ can be realized with a circularly-polarized
laser beam at the 1 Watt level, tuned near the $D_{1}$ resonance-line
and covering the entire miniature cell. High spin polarization $\left|\left\langle \mathbf{S}\right\rangle \right|=\frac{1}{2}R_{\text{P}}/\left(R_{\text{P}}+R_{\text{SD}}\right)$
could be reached, even in the presence of a small molecular background
that will be pumped through chemical-exchange collisions \citep{pumping-molecules}.
$R_{\text{P}}$ can be experimentally varied (e.g., by detuning the
pumping light from resonance) to verify the theoretical dependence
on the spin polarization $\left|\left\langle \mathbf{F}\right\rangle \right|$.
The magnetic field should be either zeroed or aligned with the optical-pumping
axis for both efficient pumping and zeroing of the Zeeman coherences.
Initial excitation of the hyperfine coherence, in low magnetic fields,
can be realized by application of a magnetic field pulse which rotates
the electron spin with little direct effect on the nuclear spin (see
SI \citep{supplemnetary-1}).  The spins can be monitored using standard
schemes (e.g., absorption spectroscopy or off-resonant Faraday rotation)
using fast photo-diodes, as the susceptibility of the vapor strongly
depends on the hyperfine coherence \citep{Happer-1970}. Fast optical
modulators \citep{Jenoptik} can be used to switch off the optical
pump beam, eliminating pump-induced relaxation during the measurement.

In conclusion, we have shown that at high spin-exchange rates, the
oscillation frequency of the hyperfine coherence is no longer constant.
Instead, many-body interactions govern the dynamics of the spins,
resulting with a collectively synchronized and surprisingly coherent
spin state. Operation at high alkali densities along with maturity
of miniaturized high-temperature cells could lead to the emergence
of highly-sensitive or highly-nonlinear applications in small-scale
devices. These include, for example, miniature SERF magnetometers
for geomagnetic fields and potentially new applications of multi-photon
processes such as coherent population trapping.

\onecolumngrid \appendix 
\setcounter{equation}{0}
\setcounter{figure}{0}
\setcounter{table}{0}
\setcounter{page}{1}
\makeatletter
\renewcommand{\theequation}{S\arabic{equation}}
\renewcommand{\thefigure}{S\arabic{figure}}
\renewcommand{\bibnumfmt}[1]{[S#1]}
\renewcommand{\citenumfont}[1]{S#1} 

\part*{Supplementary Information for ``Synchronization of strongly interacting
alkali-metal spins''}

\section{Derivation of the many-body Master equations}

The dynamics of dense thermal alkali spins is usually described by
a \textit{mean} density matrix $\bar{\rho}$ satisfying the Liouville
equation \citep{Happer-Book,Schwinger-average-density-matrix}. This
evolution yields the average spin-properties of the gas. Including
the spin-exchange interaction, this equation is given by (see Eq.~(10.20)
in \citep{Happer-Book})
\begin{equation}
\partial_{t}\bar{\rho}=-\frac{i}{\hbar}\left[H_{0},\bar{\rho}\right]+\Gamma\left\langle \mathcal{S}_{c}\bar{\rho}\mathcal{S}_{c}^{\dagger}-\bar{\rho}\right\rangle _{c},\label{eq:MEAN DENSITY MATRIX}
\end{equation}
where $H_{0}$ is the single-atom Hyperfine interaction Hamiltonian,
and $\mathcal{S}_{c}$ is the alkali-alkali scattering matrix for
a specific collision event, characterized with a particular set of
collisional parameters (including the impact parameter, the orbital
plane, and the instantaneous velocity) which are labeled with a subscript
'$c$'. $\Gamma$ is the mean collision rate and $\left\langle \cdots\right\rangle _{c}$
denotes an ensemble-average over the possible collisional realizations. 

Here we generalize this equation to describe the many-body dynamics
of $N\ggg1$ different spins, which would finally yield Eqs.~(1-3)
of the main text. We define $\rho$ as the global density matrix of
the vapor, describing the state of the $N$ electronic and $N$ nuclear
spins in the electronic ground state. Spin-exchange collisions of
alkali atoms are binary and sudden \citep{Happer-Book}, such that
after a collisional event $c$ between the $m^{\mbox{th}}$ and $n^{\mbox{th}}$
atoms, the density matrix evolves as $\rho\rightarrow\mathcal{S}_{c}^{\left(mn\right)}\rho\mathcal{S}_{c}^{\left(mn\right)\dagger}$
where $\mathcal{S}_{c}^{\left(mn\right)}$ is the scattering matrix
of the $c$ collisional event, operating on the bipartite state of
the density matrix within the $m^{\mbox{th}}$ and $n^{\mbox{th}}$
atomic subspace. On average, the many-body density matrix of the spins
$\rho$ would evolve as
\begin{equation}
\rho\left(t+dt\right)=-\frac{i}{\hbar}\left[\mathcal{H}_{0},\rho\right]dt+\sum_{m,n}\sum_{c}p_{c}^{mn}\left(dt\right)\mathcal{S}_{c}^{\left(mn\right)}\rho\mathcal{S}_{c}^{\left(mn\right)\dagger}+\left(1-p_{c}^{mn}\left(dt\right)\right)\rho\left(t\right).\label{eq:Densty}
\end{equation}
Here the first term describes the unitary evolution of the spins with
$\mathcal{H}_{0}=\hbar\sum_{n}\omega_{n}\mathbf{I}_{n}\cdot\mathbf{S}_{n}$
being the hyperfine Hamiltonian of all particles. The second term
describes the collisional interaction between the particles: $p_{c}^{mn}\left(dt\right)$
is the probability that a specific pair of atoms $m$ and $n$ had
collided during a time interval $dt$ where $c$ labels a set of specific
collision parameters. $p_{c}^{mn}\left(dt\right)$ is determined by
the kinetic theory of thermal atoms, and on average has a memory-less
time dependence (see chapter 12 in \citep{Reif}) such that $p_{c}^{mn}\left(dt\right)=\left(1-\exp\left(-\Gamma dt\right)\right)\tilde{p}_{c}^{mn}\approx\tilde{p}_{c}^{mn}\Gamma dt$,
where $\Gamma$ is the hard-sphere collision rate and $\tilde{p}_{c}^{mn}$
depends on the relative distance and velocity of the two atoms and
is nonzero when the atoms are close to each other (on the order of
the mean free path). We then find the Liouville equation
\begin{equation}
\partial_{t}\rho=-\frac{i}{\hbar}\left[\mathcal{H}_{0},\rho\right]+\Gamma\sum_{m,n}\sum_{c}\tilde{p}_{c}^{mn}\left(\mathcal{S}_{c}^{\left(mn\right)}\rho\mathcal{S}_{c}^{\left(mn\right)\dagger}-\rho\right),\label{eq:global Liouville equation scattering matrix}
\end{equation}
describing the state of the vapor for times shorter than other relaxation
rates and spatial diffusion (see section \ref{sec:Spin-relaxation-mechanisms}).
The collisional scattering matrix associated with strong spin-exchange
collisions is manifested as a correlated two-spin rotation $\mathcal{S}_{c}^{\left(mn\right)}=\exp\left(i\delta_{c}\Pi_{mn}^{e}\right)=\cos\left(\delta_{c}\right)+i\sin\left(\delta_{c}\right)\Pi_{mn}^{e}$,
where $\Pi_{mn}^{e}=\frac{1}{2}+2\mathbf{S}_{n}\cdot\mathbf{S}_{m}$
is the exchange operator of the $m-n$ spin pair ,and $\delta_{c}$
is the phase accumulated during the specific collisional event (see
Eq.~(10.252) in \citep{Happer-Book}). Substitution of this scattering
matrix in Eq.~(\ref{eq:global Liouville equation scattering matrix})
gives 
\begin{equation}
\partial_{t}\rho=-\frac{i}{\hbar}\left[\mathcal{H}_{0},\rho\right]+\Gamma\sum_{m,n}\sum_{c}\tilde{p}_{c}^{mn}\left(\sin^{2}\left(\delta_{c}\right)\left(\Pi_{mn}^{e}\rho\Pi_{mn}^{e}-\rho\right)+\frac{i}{2}\sin\left(2\delta_{c}\right)\left[\Pi_{mn}^{e},\rho\right]\right),\label{eq:fff}
\end{equation}
where the first collisional term describes real exchange of the two
spins and the second term describes collision-induced frequency shifts.
The phases $\delta_{c}$ can be estimated with either a partial-wave
analysis or using a classical path analysis \citep{happer-SERF-1977-1}.
Upon ensemble averaging, we obtain the simpler equation
\begin{equation}
\partial_{t}\rho=-\frac{i}{\hbar}\left[\mathcal{H}_{0},\rho\right]+\sum_{m,n}\Gamma_{mn}\left(\Pi_{mn}^{e}\rho\Pi_{mn}^{e}-\rho\right)\label{eq:Spin exchange equation Krauss operator}
\end{equation}
where $\Gamma_{mn}\equiv\left\langle \Gamma\sum_{c}\tilde{p}_{c}^{mn}\sin^{2}\left(\delta_{c}\right)\right\rangle _{c}$
is the average spin-exchange rate of the atomic pair $m-n$. The frequency-shift
term is omitted, since $\delta_{c}\apprge\pi$ such that, upon ensemble
averaging, $\left\langle \sum_{c}\tilde{p}_{c}^{mn}\sin\left(2\delta_{c}\right)\right\rangle _{c}$
is negligible (see Fig. 10.8 in \citep{Happer-Book}). Direct substitution
of the exchange operator $\Pi_{mn}^{e}=\frac{1}{2}+2\mathbf{S}_{n}\cdot\mathbf{S}_{m}$
results with the generalized evolution equation
\begin{equation}
\partial_{t}\rho=-i\sum_{n}\omega_{n}\left[\mathbf{I}_{n}\mathbf{S}_{n},\rho\right]+\sum_{m,n}\Gamma_{mn}\left(-\frac{3}{4}\rho+\mathbf{S}_{n}\mathbf{S}_{m}\rho+\rho\mathbf{S}_{n}\mathbf{S}_{m}+4\mathbf{S}_{n}\mathbf{S}_{m}\rho\mathbf{S}_{n}\mathbf{S}_{m}\right).\label{eq:Spin exchange liouville equation}
\end{equation}

We now assume that the quantum-correlations developed between different
colliding atoms during the interactions are raipdly lost. These coherences
are assumed to be lost for time scales longer than the short collision
duration (a few picoseconds) due to the randomness of the collision
parameters and the random choice of colliding paisr (see both Eq.~(10.105)
in \citep{Happer-Book} and the discussion in IV.D.4 in \citep{Cohen-tanoudji-book}).
We therefore consider the case that the density matrix is inter-atomic
separable and assume the simple form 
\[
\rho=\rho_{1}\otimes\ldots\rho_{n}\ldots\otimes\rho_{N},
\]
where $\rho_{n}$ is the reduced density matrix of the $n^{\mbox{th}}$
atom. Using this form, we derive the equation of motion for $\rho_{n}=\mbox{Tr}_{\neq n}\left(\rho_{n}\right)$
by partial-tracing the state of all spins but $n$, yielding 
\begin{align}
\partial_{t}\rho_{n} & =-i\omega_{n}\left[\mathbf{I}_{n}\cdot\mathbf{S}_{n},\rho_{n}\right]+\sum_{m=1}^{N}\Gamma_{mn}\left\{ -\frac{3}{4}\rho_{n}+\sum_{i=1}^{3}\left[S_{n}^{i}\rho_{n}S_{n}^{i}+\left\langle S_{m}^{i}\right\rangle \left(S_{n}^{i}\rho_{n}+\rho_{n}S_{n}^{i}\right)-2i\sum_{j=1}^{3}\epsilon_{ijk}\left\langle S_{m}^{k}\right\rangle S_{n}^{i}\rho_{n}S_{n}^{j}\right]\right\} \nonumber \\
= & -i\omega_{n}\left[\mathbf{I}_{n}\cdot\mathbf{S}_{n},\rho_{n}\right]+\sum_{m}\Gamma_{mn}\left(-\frac{3}{4}\rho_{n}+\mathbf{S}_{n}\rho_{n}\mathbf{S}_{n}+\left\langle \mathbf{S}_{m}\right\rangle (\rho_{n}\mathbf{S}_{n}+\mathbf{S}_{n}\rho_{n}-2i\mathbf{S}_{n}\times\rho_{n}\mathbf{S}_{n})\right),\label{eq: reduced Density matrix Liouville equation}
\end{align}
where $\left\langle \mathbf{S}_{m}\right\rangle \equiv\mbox{Tr}\left(\rho_{m}\mathbf{S}_{m}\right)$
is the mean electronic spin of the $m^{\mbox{th}}$ atom, and $\epsilon_{ijk}$
is Levi-Civita symbol. Equation~(\ref{eq: reduced Density matrix Liouville equation})
is the many-body generalization for the mean-field evolution of the
spin-exchange interaction (see \citep{Happer-1972}, in particular
Eqs.~(VI.8) and (VI.15)).

\section{Derivation of the many-body Bloch equations}

The total evolution of the reduced density matrix in Eq.~(\ref{eq: reduced Density matrix Liouville equation})
can be decomposed into the following terms: 
\[
\frac{d}{dt}\rho_{n}=\underset{\mbox{hyperfine}}{\underbrace{-i\omega_{n}\left[\mathbf{I}_{n}\cdot\mathbf{S}_{n},\rho_{n}\right]}}\underset{\mbox{SE}_{1}}{\underbrace{-(\sum_{m\neq n}^{N}\Gamma_{mn})(\frac{3}{4}\rho_{n}-\mathbf{S}_{n}\rho_{n}\mathbf{S}_{n})}}+\underset{\mbox{SE}_{2}}{\underbrace{(\sum_{m\neq n}^{N}\Gamma_{mn}\left\langle \mathbf{S}_{m}\right\rangle )(\rho_{n}\mathbf{S}_{n}+\mathbf{S}_{n}\rho_{n}-2i\mathbf{S}_{n}\times\rho_{n}\mathbf{S}_{n})}}.
\]
The first term is the hyperfine coupling. The second and third terms
are respectively linear and nonlinear, and they account respectively
for the destructive and conservative parts of the spin-exchange interaction.
We shall examine the evolution of the different moments $\left\langle S_{n}^{i}\right\rangle $,
$\left\langle I_{n}^{i}\right\rangle $, and $\left\langle A_{n}^{i}\right\rangle $,
utilizing the commutation relations of the electronic spins $\left\{ S_{m}^{i},S_{m}^{j}\right\} =\frac{1}{2}\delta_{ij}$
and $\left[S_{m}^{i},S_{n}^{j}\right]=i\delta_{mn}\epsilon_{ijk}S_{k}$
and the nuclear spins $\left\{ I_{m}^{i},I_{m}^{j}\right\} =\frac{1}{2}\delta_{ij}$
and $\left[I_{m}^{i},I_{n}^{j}\right]=i\delta_{mn}\epsilon_{ijk}I_{k}$.
The evolution due to the \textbf{\uline{hyperfine}} coupling is
given by

{\small{}
\begin{align*}
\frac{d}{dt}\left(\left\langle S_{n}^{i}\right\rangle \right)_{{\scriptscriptstyle \textsc{\textnormal{hpf}}}} & =-i\omega_{n}\sum_{j}\mbox{Tr}(S_{n}^{i}\left[I_{n}^{j}S_{n}^{j},\rho_{n}\right])=-i\omega_{n}\sum_{j}\mbox{Tr}(S_{n}^{i}S_{n}^{j}I_{n}^{j}\rho_{n}-\rho_{n}I_{n}^{j}S_{n}^{j}S_{n}^{i})\\
 & =-i\omega_{n}\sum_{j}\mbox{Tr}((\tfrac{1}{4}\delta_{ij}+\tfrac{i}{2}\epsilon_{ijk}S_{n}^{k})I_{n}^{j}\rho_{n}-\rho_{n}I_{n}^{j}(\tfrac{1}{4}\delta_{ji}+\tfrac{i}{2}\epsilon_{jik}S_{n}^{k}))=\omega_{n}\sum_{j}\epsilon_{ijk}\left\langle I_{n}^{j}S_{n}^{k}\right\rangle =\omega_{n}\left\langle A_{n}^{i}\right\rangle ,\\
\frac{d}{dt}\left(\left\langle I_{n}^{i}\right\rangle \right)_{{\scriptscriptstyle \textsc{\textnormal{hpf}}}} & =-i\omega_{n}\sum_{j}\mbox{Tr}(I_{n}^{i}\left[I_{n}^{j}S_{n}^{j},\rho_{n}\right])=\omega_{n}\sum_{j}\epsilon_{ijk}\mbox{Tr}(S_{n}^{j}I_{n}{}^{k})=-\omega_{n}\left\langle A_{n}^{i}\right\rangle ,\\
\frac{d}{dt}\left(\left\langle A_{n}^{i}\right\rangle \right)_{\mbox{\ensuremath{{\scriptscriptstyle \textsc{\textnormal{hpf}}}}}} & =-i\omega_{n}\sum_{mjk}\epsilon_{ijk}\mbox{Tr}(S_{n}^{j}I_{n}^{k}\left[I_{n}^{m}S_{n}^{m},\rho_{n}\right])\\
 & =-i\omega_{n}\sum_{mjk}\epsilon_{ijk}\mbox{Tr}[(\tfrac{1}{4}\delta_{jm}+\tfrac{i}{2}\epsilon_{jmq}S_{n}^{q})(\tfrac{1}{4}\delta_{km}+\tfrac{i}{2}\epsilon_{kmp}I_{n}^{p})\rho_{n}-\rho_{n}(\tfrac{1}{4}\delta_{mk}+\tfrac{i}{2}\epsilon_{mkp}I_{n}^{p})(\tfrac{1}{4}\delta_{mj}+\tfrac{i}{2}\epsilon_{mjq}S_{n}^{q})]\\
 & =\frac{\omega_{n}}{4}\sum_{jk}\epsilon_{ijk}\mbox{Tr}(\epsilon_{kjp}I_{n}^{p}\rho_{n}+\epsilon_{jkq}S_{n}^{q}\rho_{n})=\frac{\omega_{n}}{4}\sum_{jk}\mbox{Tr}(-2\delta_{ip}I_{n}^{p}\rho_{n}+2\delta_{iq}S_{n}^{q}\rho_{n})=\frac{\omega_{n}}{2}(\left\langle S_{n}^{i}\right\rangle -\left\langle I_{n}^{i}\right\rangle ).
\end{align*}
}The evolution due to the \textbf{\uline{linear spin-exchange term}}
($\mbox{SE}_{1}$) is given by{\small{}
\begin{align*}
\frac{d}{dt}\left(\left\langle S_{n}^{i}\right\rangle \right)_{{\scriptscriptstyle \textsc{\textnormal{SE}}}_{1}} & =-\frac{3}{4}\sum_{m}\Gamma_{mn}\left\langle S_{n}^{i}\right\rangle +\sum_{m}\Gamma_{mn}\sum_{j}\mbox{Tr}(\rho_{n}S_{n}^{j}S_{n}^{i}S_{n}^{j})=-\sum_{m}\Gamma_{mn}\left\langle S_{n}^{i}\right\rangle ,\\
\frac{d}{dt}\left(\left\langle I_{n}^{i}\right\rangle \right)_{{\scriptscriptstyle \textsc{\textnormal{SE}}}_{1}} & =-\sum_{m}\Gamma_{mn}\mbox{Tr}[I_{n}^{i}(\tfrac{3}{4}\rho_{n}-\sum_{j}S_{n}^{j}\rho_{n}S_{n}^{j})]-\tfrac{3}{4}\sum_{m}\Gamma_{mn}\left\langle I_{n}^{i}\right\rangle +\tfrac{3}{4}\sum_{m}\Gamma_{mn}\left\langle I_{n}^{i}\right\rangle =0,\\
\frac{d}{dt}\left(\left\langle A_{n}^{i}\right\rangle \right)_{{\scriptscriptstyle \textsc{\textnormal{SE}}}_{1}} & =-\sum_{m}\Gamma_{mn}\mbox{Tr}[\sum_{jk}\epsilon_{ijk}S_{n}^{j}I_{n}^{k}(\tfrac{3}{4}\rho_{n}-\sum_{q}S_{n}^{q}\rho_{n}S_{n}^{q})]\\
 & =-\frac{3}{4}\sum_{m}\Gamma_{mn}\left\langle A_{n}^{i}\right\rangle +\sum_{m}\Gamma_{mn}\sum_{jkq}\epsilon_{ijk}\left\langle S_{n}^{q}S_{n}^{j}S_{n}^{q}I_{n}^{k}\right\rangle \\
 & =-\frac{3}{4}\sum_{m}\Gamma_{mn}\left\langle A^{i}\right\rangle -\frac{1}{4}\sum_{m}\Gamma_{mn}\sum_{jk}\epsilon_{ijk}\left\langle S^{j}I^{k}\right\rangle =-\sum_{m}\Gamma_{mn}\left\langle A_{n}^{i}\right\rangle ,
\end{align*}
}where in the last equations, we used the identities $S_{n}^{j}S_{n}^{i}S_{n}^{j}=-\frac{1}{4}S_{n}^{j}$
and $\sum_{qp}\epsilon_{qpl}S_{n}^{q}S_{n}^{j}S_{n}^{p}=-\frac{i}{4}\delta_{jl}$.
The evolution due to the \textbf{\uline{nonlinear spin-exchange
term}} $(\mbox{SE}_{2}$) is given by

{\small{}
\begin{align*}
\frac{d}{dt}\left(\left\langle S_{n}^{i}\right\rangle \right)_{{\scriptscriptstyle \textsc{\textnormal{SE}}}_{2}} & =\sum_{m}\Gamma_{nm}\sum_{l}\left\langle S_{m}^{l}\right\rangle \mbox{Tr}[S_{n}^{i}(\rho_{n}S_{n}^{l}+S_{n}^{l}\rho_{n}-2i\sum_{jk}\epsilon_{jkl}S_{n}^{j}\rho_{n}S_{n}^{k})]\\
 & =\sum_{m}\Gamma_{nm}\sum_{l}\left\langle S_{m}^{l}\right\rangle \mbox{Tr}[\rho_{n}(\tfrac{1}{4}\delta_{li}+\tfrac{i}{2}\epsilon_{lip}S_{n}^{p})+(\tfrac{1}{4}\delta_{il}+\tfrac{i}{2}\epsilon_{ilp}S_{n}^{p})\rho_{n}+2i\sum_{jk}\epsilon_{kjl}S_{n}^{k}S_{n}^{i}S_{n}^{j}\rho_{n}]\\
 & =\frac{1}{2}\sum_{m}\Gamma_{nm}\left\langle S_{m}^{i}\right\rangle -2i\sum_{m}\Gamma_{nm}\sum_{l}\left\langle S_{m}^{l}\right\rangle \frac{i}{4}\delta_{il}=\sum_{m}\Gamma_{nm}\left\langle S_{m}^{i}\right\rangle ,\\
\frac{d}{dt}\left(\left\langle I_{n}^{i}\right\rangle \right)_{{\scriptscriptstyle \textsc{\textnormal{SE}}}_{2}} & =\sum_{m}\Gamma_{nm}\sum_{l}\left\langle S_{m}^{l}\right\rangle \mbox{Tr}[I_{n}^{i}(\rho_{n}S_{n}^{l}+S_{n}^{l}\rho_{n}-2i\sum_{jk}\epsilon_{jkl}S_{n}^{j}\rho_{n}S_{n}^{k})]\\
 & =\sum_{m}\Gamma_{nm}\sum_{l}\left\langle S_{m}^{l}\right\rangle \{\left\langle I_{n}^{i}S_{n}^{l}\right\rangle -2i\sum_{jk}\epsilon_{jkl}\mbox{Tr}[\rho_{n}I_{n}^{i}(\tfrac{1}{4}\delta_{kj}+\tfrac{i}{2}\epsilon_{kjp}S_{n}^{p})]\}=0,\\
\frac{d}{dt}\left(\left\langle A_{n}^{i}\right\rangle \right)_{{\scriptscriptstyle \textsc{\textnormal{SE}}}_{2}} & =\sum_{m}\Gamma_{nm}\sum_{jkl}\epsilon_{ijk}\left\langle S_{m}^{l}\right\rangle \mbox{Tr}[S_{n}^{j}I_{n}^{k}(\rho_{n}S_{n}^{l}+S_{n}^{l}\rho_{n}-2i\sum_{pq}\epsilon_{pql}S_{n}^{p}\rho_{n}S_{n}^{q})]\\
 & =\sum_{m}\Gamma_{nm}\sum_{jkl}\epsilon_{ijk}\left\langle S_{m}^{l}\right\rangle \mbox{Tr}[I_{n}^{k}\rho_{n}(\tfrac{1}{4}\delta_{lj}+\tfrac{i}{2}\epsilon_{ljp}S_{n}^{p})+\rho_{n}I_{n}^{k}(\tfrac{1}{4}\delta_{jl}+\tfrac{i}{2}\epsilon_{jlp}S_{n}^{p})-2i\sum_{pq}\epsilon_{pql}S_{n}^{q}S_{n}^{j}S_{n}^{p}I_{n}^{k}\rho_{n}]\\
 & =\frac{1}{2}\sum_{m}\Gamma_{nm}\sum_{jkl}\epsilon_{ijk}\left\langle S_{m}^{l}\right\rangle \left\langle I_{n}^{k}\right\rangle -2i\sum_{m}\Gamma_{nm}\sum_{jkl}\epsilon_{ijk}\left\langle S_{m}^{l}\right\rangle \frac{i}{4}\delta_{jl}\left\langle I^{k}\right\rangle =\sum_{m}\Gamma_{nm}\sum_{jk}\epsilon_{ijk}\left\langle S_{m}^{j}\right\rangle \left\langle I_{n}^{k}\right\rangle \\
 & =\sum_{m}\Gamma_{nm}\left(\left\langle \mathbf{S}_{m}\right\rangle \times\left\langle \mathbf{I}_{m}\right\rangle \right)_{i}.
\end{align*}
}Combining the above 9 terms, we arrive at the Bloch Eqs.~(1-3) of
the main text:

\begin{align}
\frac{d}{dt}\left\langle \mathbf{S}_{n}\right\rangle \, & =\omega_{n}\left\langle \mathbf{A}_{n}\right\rangle +\sum_{m}\Gamma_{mn}\left(\left\langle \mathbf{S}_{m}\right\rangle -\left\langle \mathbf{S}_{n}\right\rangle \right)\label{eq:S dynamics many body-1-1}\\
\frac{d}{dt}\left\langle \mathbf{I}_{n}\right\rangle \, & =-\omega_{n}\left\langle \mathbf{A}_{n}\right\rangle \label{eq: I dynamics many body-1-1}\\
\frac{d}{dt}\left\langle \mathbf{A}_{n}\right\rangle  & =-\frac{\omega_{n}}{2}\left(\left\langle \mathbf{S}_{n}\right\rangle -\left\langle \mathbf{I}_{n}\right\rangle \right)-\sum_{m}\Gamma_{mn}\left\langle \mathbf{A}_{n}\right\rangle +\sum_{m}\Gamma_{mn}\left\langle \mathbf{S}_{m}\right\rangle \times\left\langle \mathbf{I}_{n}\right\rangle .\label{eq:N dynamics many body I,S-1-1}
\end{align}

\section{Approximations in the strong-interaction regime}

In this part, we derive Eqs.~(9-10) in the main text, which approximate
the dynamics of the vapor in the strong interaction regime $\sum_{m}\Gamma_{mn}\gg\omega_{n}$.
We first transform the first-order differential equations (\ref{eq:S dynamics many body-1-1}-\ref{eq:N dynamics many body I,S-1-1})
into second-order differential equations by eliminating the torque
observable $\left\langle \mathbf{A}_{n}\right\rangle $

\begin{align}
\frac{d}{dt}\left\langle \mathbf{S}_{n}\right\rangle  & =-\frac{d}{dt}\left\langle \mathbf{I}_{n}\right\rangle +\sum_{m}\Gamma_{mn}\left(\left\langle \mathbf{S}_{m}\right\rangle -\left\langle \mathbf{S}_{n}\right\rangle \right)\label{eq: tmp eq S}\\
\frac{d^{2}}{dt^{2}}\left\langle \mathbf{I}_{n}\right\rangle  & +\sum_{m}\Gamma_{mn}\frac{d}{dt}\left\langle \mathbf{I}_{n}\right\rangle +\frac{\omega_{n}^{2}}{2}\left(\left\langle \mathbf{I}_{n}\right\rangle -\left\langle \mathbf{S}_{n}\right\rangle \right)+\omega_{n}\sum_{m}\Gamma_{mn}\left\langle \mathbf{S}_{m}\right\rangle \times\left\langle \mathbf{I}_{n}\right\rangle =0.\label{eq:tmp eq I}
\end{align}
Eq.~(\ref{eq: tmp eq S}) can be used to derive Eq.~(9) in the main
text, which describes the dynamics of the total spins $\left\langle \mathbf{F}_{n}\right\rangle =\left\langle \mathbf{S}_{n}\right\rangle +\left\langle \mathbf{I}_{n}\right\rangle $,
\begin{equation}
\frac{d}{dt}\left\langle \mathbf{F}_{n}\right\rangle =\sum_{m}\Gamma_{mn}\left(\left\langle \mathbf{S}_{m}\right\rangle -\left\langle \mathbf{S}_{n}\right\rangle \right).\label{eq: kuramoto1}
\end{equation}
We now rewrite Eq.~(\ref{eq:tmp eq I}) in terms of the total spins

\[
\frac{d^{2}}{dt^{2}}\left\langle \mathbf{F}_{n}\right\rangle -\frac{d^{2}}{dt^{2}}\left\langle \mathbf{S}_{n}\right\rangle =\omega_{n}^{2}\left(\left\langle \mathbf{S}_{n}\right\rangle -\nicefrac{1}{2}\left\langle \mathbf{F}_{n}\right\rangle \right)-\sum_{m}\Gamma_{mn}\frac{d}{dt}\left(\left\langle \mathbf{F}_{n}\right\rangle -\left\langle \mathbf{S}_{n}\right\rangle \right)+-\omega_{n}\sum_{m}\Gamma_{mn}\left\langle \mathbf{S}_{m}\right\rangle \times\left\langle \mathbf{F}_{n}\right\rangle +\omega_{n}\sum_{m}\Gamma_{mn}\left\langle \mathbf{S}_{m}\right\rangle \times\left\langle \mathbf{S}_{n}\right\rangle .
\]
In the strong-interaction regime, the oscillations slow down due to
motional narrowing, rendering the second-order derivatives on the
left-hand side negligible. We furthermore neglect the last term on
the right-hand side, as $\left\langle \mathbf{S}_{m}\right\rangle \approx\left\langle \mathbf{S}_{n}\right\rangle $
due to the synchronization of the spins. The equation thus simplifies
to a first-order differential equation

\begin{align*}
\sum_{m}\Gamma_{mn}\frac{d}{dt}\left\langle \mathbf{S}_{n}\right\rangle  & =-\omega_{n}^{2}\left(\left\langle \mathbf{S}_{n}\right\rangle -\nicefrac{1}{2}\left\langle \mathbf{F}_{n}\right\rangle \right)+\sum_{m}\Gamma_{mn}\frac{d}{dt}\left(\left\langle \mathbf{F}_{n}\right\rangle \right)+\omega_{n}\sum_{m}\Gamma_{mn}\left\langle \mathbf{S}_{m}\right\rangle \times\left\langle \mathbf{F}_{n}\right\rangle .
\end{align*}
Finally, defining the mean relaxation of the $n\mbox{\ensuremath{^{th}}}$
atom as $\Gamma_{n}\equiv\sum_{m}\Gamma_{mn}$ and substituting Eq.~(\ref{eq: kuramoto1}),
we obtain Eq.~(10) of the main text

\begin{align*}
\frac{d}{dt}\left\langle \mathbf{S}_{n}\right\rangle  & =\sum_{m}\Gamma_{mn}\left(\left\langle \mathbf{S}_{m}\right\rangle -\left\langle \mathbf{S}_{n}\right\rangle \right)+\frac{\omega_{n}}{\Gamma_{n}}\sum_{m}\Gamma_{mn}\left\langle \mathbf{S}_{m}\right\rangle \times\left\langle \mathbf{F}_{n}\right\rangle -\frac{\omega_{n}^{2}}{\Gamma_{n}}\left(\left\langle \mathbf{S}_{n}\right\rangle -\nicefrac{1}{2}\left\langle \mathbf{F}_{n}\right\rangle \right)\\
= & \omega_{n}\left\langle \mathbf{S}_{n}\right\rangle \times\left\langle \mathbf{F}_{n}\right\rangle +\left(\sum_{m}\Gamma_{mn}\left(\left\langle \mathbf{S}_{m}\right\rangle -\left\langle \mathbf{S}_{n}\right\rangle \right)\right)\left(1+\times\left\langle \mathbf{F}_{n}\right\rangle \frac{\omega_{n}}{\Gamma_{n}}\right)-\frac{\omega_{n}^{2}}{\Gamma_{n}}\left(\left\langle \mathbf{S}_{n}\right\rangle -\nicefrac{1}{2}\left\langle \mathbf{F}_{n}\right\rangle \right)\\
= & \omega_{n}\left\langle \mathbf{S}_{n}\right\rangle \times\left\langle \mathbf{F}_{n}\right\rangle +\sum_{m}\Gamma_{mn}\left(\left\langle \mathbf{S}_{m}\right\rangle -\left\langle \mathbf{S}_{n}\right\rangle \right)-\frac{\omega_{n}^{2}}{\Gamma_{n}}\left(\left\langle \mathbf{S}_{n}\right\rangle -\nicefrac{1}{2}\left\langle \mathbf{F}_{n}\right\rangle \right),
\end{align*}
where, in the last equality, we neglected a high order term involving
$\times\left\langle \mathbf{F}_{n}\right\rangle \frac{\omega_{n}}{\Gamma_{n}}$. 

\section{Experimental Roadmap: Spin relaxation mechanisms and initialization
of hyperfine coherence\label{sec:Spin-relaxation-mechanisms}}

The dominant spin-relaxation mechanisms in the high-temperature atomic
vapor we consider are \citep{Walker-2000-relaxations-singlet-1}:
a. interaction with the walls at a rate $R_{\text{wall}}$. b. $\text{K-K}$
destructive collisions at a rate $R_{\text{KK}}$. c. Molecular relaxation
by singlet dimers $^{1}\Sigma_{g}^{+}$ at a rate $R_{\text{S}}$.
d. Spin rotation through collisions with $\text{\ensuremath{N_{2}}}$
at a rate $R_{\text{buff}}$. Other relaxation mechanisms, such as
magnetic fields gradients \citep{Happer-Romalis-SEOP-1998-1} can
be made small. The total electronic relaxation rate is then given
by 
\[
R_{\text{SD}}=R_{\text{wall}}+R_{\text{KK}}+R_{\text{S}}+R_{\text{buff}}.
\]
We estimate the electronic relaxation $R_{\text{wall}}$ by assuming
that the walls are completely depolarizing and consider the least
decaying diffusion mode (see Eq. (10.286) in \citep{Happer-Book})
\[
R_{\text{wall}}\approx4\pi^{2}QD/L^{2}\approx10^{6}\;\text{se\ensuremath{c^{-1}}},
\]
where $D\approx0.4\;\text{c\ensuremath{m^{2}}/sec}$ is the diffusion
coefficient for $1\;\text{amg}$ of $\text{\ensuremath{N_{2}}}$ and
$Q=6$ is the slowing down factor (for $I=3/2$) accounting for the
loss of nuclear spin during the interaction with the wall, see Eq.
(10.271) in Ref.~\citep{Happer-Book}. Spin destruction of alkali-alkali
collisions consists of two main mechanisms: spin rotation in binary
collisions and spin-axis relaxation in molecular triplet dimers \citep{Walker-2000-spin-destruction-triplet,Walker-spin-relaxation-theory}.
These two interactions were found to have equal magnitudes and together
destruct the spin at a rate 
\[
R_{\text{KK}}=n_{{\scriptscriptstyle \textsc{\textnormal{K}}}}\sigma_{\text{KK}}\bar{v}\approx2\cdot10^{5}\;\text{se\ensuremath{c^{-1}}}
\]
where we used $n_{{\scriptscriptstyle \textsc{\textnormal{K}}}}=1.7\cdot10^{18}\;\text{c\ensuremath{m^{-3}}},$
$\bar{v}\approx10^{5}\;\mbox{cm/sec}$ and we assumed the cross-section
$\sigma_{\text{KK}}=10^{-18}\;\text{c\ensuremath{m^{-2}}}$, which
was measured at low temperatures \citep{Walker-Potassium-SD-Crosssection},
with no known dependence on temperature variation. The current theoretical
models predict an order of magnitude smaller value for the $\sigma_{\text{KK}}$
we use \citep{Walker-2000-spin-destruction-triplet,Walker-spin-relaxation-theory},
and this cross section should be considered only as an order of magnitude
estimate. To validate the molecular estimation at higher temperatures,
we also compute the chemical potential for triplet dimers at $T=620^{\circ}\;\text{C }$by
following a procedure similar to Ref.~\citep{Walker-2000-relaxations-singlet-1}
and using the molecular potential in \citep{Krauss-Stevens-potential}.
We then estimate that the chemical equilibrium coefficient of the
triplet dimers is $\mathcal{K}_{T}=3\cdot10^{-23}\;\text{c\ensuremath{m^{3}}}$,
using a triplet binding energy of $D_{e}^{\left(T\right)}=0.032\;\text{eV}<k_{B}T$
and assuming that during a molecular lifetime the spin loses a fraction
$\alpha_{T}\leq1$ of its coherence. The estimated triplet destruction
at $T=620\;\text{\ensuremath{^{\circ}}C}$ is then bounded by $R_{\text{KK}}\approx\alpha_{T}\tau_{c}^{-1}\mathcal{K}_{T}n_{{\scriptscriptstyle \textsc{\textnormal{K}}}}<3\cdot10^{6}\;\text{sec}^{-1}$,
where $\tau_{c}^{-1}=n_{{\scriptscriptstyle \textsc{\textnormal{N2}}}}\left(\sigma\bar{v}\right)_{{\scriptscriptstyle \textsc{\textnormal{K2-N2}}}}\approx6\cdot10^{10}\;\text{sec}^{-1}$
is the hard-sphere collision rate with $\text{\ensuremath{N_{2}}}$
molecules (which serve as third bodies).

Singlet dimers are the most populated molecular state, with estimated
dimer to monomer ratio limited to a few percents at $T=620\;\text{\ensuremath{^{\circ}}C}$
(Using measured data of the molecular partial pressure of $\text{\ensuremath{K_{2}}}$
by Ref.~\citep{Nesmeyanov} we estimate a molecular fraction of $3.5\%$,
and using the potentials of Ref.~\citep{Krauss-Stevens-potential}
we numerically calculate the chemical potential following a similar
procedure to~\citep{Walker-2000-relaxations-singlet-1} and estimate
a fraction of $5\%$. We verify that our chemical potential fits the
results of Ref.~\citep{Walker-2000-relaxations-singlet-1} at low
temperatures). We note, however, that the molecular fraction calculated
here could be larger for alkali halides, and therefore pure alkali
metal should be used instead \citep{Nesmeyanov}.  The atomic decoherence
due to singlet dimers results mainly from molecular dissociation,
where relaxation of the nuclear spins during a molecular lifetime
is found negligible. Upon dissociation of the dimer, the total spins
of the atomic pair is conserved but the atoms could possibly result
with hyperfine coherence being unsynchronized with the rest of the
atomic ensemble. Such atoms would spin-thermalize with the rest of
the ensemble and contribute to the total decoherence rate. We approximate
this rate by 
\[
R_{\text{S}}=\alpha_{S}(\frac{n_{{\scriptscriptstyle \textsc{\textnormal{K2}}}}}{n_{{\scriptscriptstyle \textsc{\textnormal{K}}}}})\cdot\tau_{c}^{-1}\exp\left(-\frac{D_{e}^{\left(S\right)}}{k_{B}T}\right)<1.5\cdot10^{6}\;\text{se\ensuremath{c^{-1}}}
\]
where $D_{e}^{\left(S\right)}\approx0.55\;\text{eV}$ is the molecular
binding energy, $n_{{\scriptscriptstyle \textsc{\textnormal{K2}}}}$
is the density of singlet dimers and $\alpha_{S}\leq1$ is the amount
of coherence lost at a single dissociation of a singlet dimer. The
singlet dimers have no electronic spin and during their lifetime only
the nuclear spin is subject to relaxation. The nuclear spin is subject
to both electric-quadruple and nuclear spin interactions \citep{Walker-2000-relaxations-singlet-1}.
As a singlet molecule experiences multiple collisions before dissociation,
the nuclear spin relaxation is given by $R_{s}^{\left(1\right)}\approx\left(\frac{2}{3}\Omega_{q}^{2}+c^{2}\left\langle J^{2}\right\rangle \right)\tau_{R}<10\;\text{se\ensuremath{c^{-1}}}$
where $\Omega_{q}\approx1.9\cdot10^{5}\;\text{se\ensuremath{c^{-1}}}$
is the quadruple interaction strength, $c\sqrt{\left\langle J^{2}\right\rangle }\approx3.5\cdot10^{4}\;\text{se\ensuremath{c^{-1}}}$
is the spin rotation interaction strength, and $\tau_{R}$ is the
typical reorienting collision time. In our setup $\tau_{R}$ is equally
split between collision with buffer gas atoms, which reorient the
molecular rotation ($J$) and chemical-exchange collisions with other
alkali atoms, which swap the nuclear spin of one of the nucleus (which
is equivalent to reorientation of the nuclear spin) such that overall
$\tau_{R}^{-1}\approx(n_{{\scriptscriptstyle \textsc{\textnormal{N2}}}}\left(\sigma_{J}\bar{v}\right))^{-1}+(n_{{\scriptscriptstyle \textsc{\textnormal{K}}}}\left(\sigma\bar{v}\right)_{{\scriptscriptstyle \textsc{\textnormal{K-K2}}}})^{-1}\approx$$6\cdot10^{9}\,\text{se\ensuremath{c^{-1}}},$
where we used the chemical-exchange rate $\left(\sigma\bar{v}\right)_{{\scriptscriptstyle \textsc{\textnormal{K-K2}}}}\approx1.5\cdot10^{-9}\;\text{c\ensuremath{m^{3}}}\text{/sec}$
and the reorientation rate $\sigma_{J}\bar{v}_{{\scriptscriptstyle \textsc{\textnormal{K2-N2}}}}\approx1.5\cdot10^{-10}\;\text{c\ensuremath{m^{3}}}\text{/sec}$
based on measurements with $\text{R\ensuremath{b_{2}}}$ dimers \citep{Walker-2000-relaxations-singlet-1}.
We note that atomic potassium encounter also frequent chemical-exchange
collisions with singlet dimers, at a rate $R_{{\scriptscriptstyle \textsc{\textnormal{CE}}}}=n_{{\scriptscriptstyle \textsc{\textnormal{K2}}}}\left(\sigma\bar{v}\right)_{{\scriptscriptstyle \textsc{\textnormal{K-K2}}}}\approx7.5\cdot10^{7}\;\text{se\ensuremath{c^{-1}}}$,
which in contrast to $R_{\text{S}}$, is not suppressed with the Boltzmann
factor $\exp(-D_{e}^{\left(S\right)}/k_{B}T)$ \citep{Chemical-exchange-Happer}.
These collisions conserve the electronic spin and can be thought of
as an exchange operation of one atomic nucleus with one of the nuclei
in a molecule. The molecular nuclei, previously formed from a pair
of atomic alkali, are oriented with almost the same direction as the
alkali one. Therefore, the chemical exchange collisions play a similar
role to atomic spin-exchange collisions, and its effect on the atomic
vapor is to increase $R_{\text{SE}}$ but not $R_{\text{SD}}$. Therefore
in the strong-interaction regime it should not impose any additional
relaxation. We note that application of high magnetic fields can significantly
suppress the dimer part of the relaxation, for both the singlet and
triplet states \citep{Walker-2000-spin-destruction-triplet-1}.

Relaxation due to collisions with buffer gas is estimated as 
\[
R_{\text{buff}}=n_{{\scriptscriptstyle \textsc{\textnormal{N2}}}}\sigma'\bar{v}'<10^{4}\;\text{se\ensuremath{c^{-1}}}
\]
 where $n_{{\scriptscriptstyle \textsc{\textnormal{N2}}}}=2.5\cdot10^{19}\;\text{c\ensuremath{m^{-3}}}$
is the nitrogen density, $\sigma'\approx10^{-21}\;\text{c\ensuremath{m^{2}}}$
is the spin-rotation cross section of $\text{K-\ensuremath{N_{2}}}$,
estimated at $T=620\;\text{\ensuremath{^{\circ}}C}$ (with the $T^{3.7}$
dependence taken into account) and $\bar{v}'\approx1.3\cdot10^{5}\;\text{cm/sec}$
is the mean thermal velocity of the $\text{K-\ensuremath{N_{2}}}$
pair \citep{Happer-Book}. In conclusion, for the experimental conditions
we outline in the main text, we predict $R_{\text{SD}}<5\cdot10^{6}\;\text{se\ensuremath{c^{-1}}}$,
such that spin-exchange is expected to be the dominant relaxation
mechanism even for very dense vapor at high temperatures.

Initial excitation of the hyperfine coherence, in low magnetic fields,
can be realized by application of a magnetic field pulse which rotates
the electron spin (which has a gyromagnetic ratio $g_{s}=2.8\,\text{MHz/G}$)
with little direct effect on the nuclear spin (which has a gyromagnetic
ratio of $g_{I}=78\,\text{Hz/G}$ for $^{41}\text{K}$ \citep{Potassium_Tiecke}).
A general pulse would excite simultaneously both Zeeman and hyperfine
coherences. It is possible however to excite a specific hyperfine
coherence magnetically while leaving the Zeeman coherence unexcited
by shaping the applied magnetic pulse. For example, if the pulsed
magnetic field is oriented perpendicular to the optical-pumping axis,
and consists of a single sine burst ($B_{\perp}\sin\left(\omega_{B}t\right)$
for $0\leq t\leq2\pi/\omega_{B}$) then it would rotate the electronic
spin back and forth. For $\omega_{B}<\omega_{{\scriptscriptstyle \textsc{\textnormal{K}}}}$
the nuclear spin is strongly coupled to the electronic spin and follows
its track such that at the end of the pulse the spins return to their
starting point, and no coherence is introduced. If $\omega_{B}>\omega_{{\scriptscriptstyle \textsc{\textnormal{K}}}}$
only the electronic spin precesses by the pulse and the hyperfine
interaction with $I$ accumulates an additional phase (azimuth $\sim\omega_{{\scriptscriptstyle \textsc{\textnormal{K}}}}/\omega_{B}$,
and elevation $\sim\frac{1}{4}g_{s}B_{\perp}/\omega_{B}$) and the
spins would not return to their initial point, exciting mainly the
$\lambda_{1}^{+}$ hyperfine coherence, while the Zeeman coherences
are zeroed at the end of the pulse. Low inductance short wires can
support GHz bandwidth pulses and can be positioned in the proximity
of the cell \citep{NV_wire}.

\end{document}